\documentclass[aip,apl,reprint]{revtex4-1}
\usepackage[pdftex]{graphicx}
\usepackage{dcolumn}
\usepackage{bm}

\begin{document}
%\draft
\title {Selective Activation of an Isolated Magnetic Skyrmion in a Ferromagnet with Microwave Electric Fields}

\author{Akihito Takeuchi}
\email{akihito@phys.aoyama.ac.jp}
\affiliation{Department of Physics and Mathematics, Aoyama Gakuin University, Sagamihara, Kanagawa 252-5258, Japan}
\author{Masahito Mochizuki}
\email{masa\_mochizuki@waseda.jp}
\affiliation{Department of Applied Physics, Waseda University, Okubo, Shinjuku-ku, Tokyo 169-8555, Japan}
\affiliation{PRESTO, Japan Science and Technology Agency, Kawaguchi, Saitama 332-0012, Japan}
\begin{abstract}
We theoretically reveal that pure eigenmodes of an isolated magnetic skyrmion embedded in a ferromagnetic environment can be selectively activated using microwave electric fields without exciting gigantic ferromagnetic resonances, in contrast to conventional methods using microwave magnetic fields. We also demonstrate that this selective activation of a skyrmion can efficiently drive its translational motion in a ferromagnetic nanotrack under application of an external magnetic field inclined from the normal direction. We find that a mode with combined breathing and rotational oscillations induces much faster skyrmion propagation than the breathing mode studied previously by Wang $et$ $al.$ [Phys. Rev. B {\bf 92}, 020403(R) (2015)].
\end{abstract}
\pacs{76.50.+g,78.20.Ls,78.20.Bh,78.70.Gq}
%%75.85.+t Magnetoelectric effects, multiferroics
%%76.50.+g FM, AM, and ferrimagnetic resonances; spin-wave resonance
%%77.80.-e Ferroelectricity and antiferroelectricity
%%78.20.Ls Magneto-optical effects
%%78.20.Bh Theory, models, and numerical simulation
%%78.70.Gq Microwave and radio-frequency interactions
\maketitle
%\sloppy \maketitle

%%%%%%%%%%%%%%%%%%%%%%%%%%%%%%%%%%%%%%%%%%%%%%%%%%%%%%%%%%%%%
\begin{figure}
\includegraphics[scale=1.0]{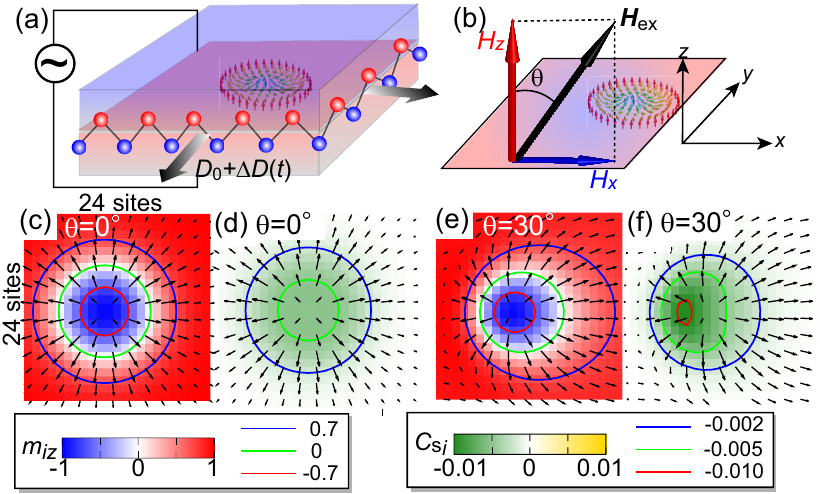}
\caption{(color online). (a) Schematics of a magnetic bilayer system hosting skyrmions stabilized by the interfacial Dzyaloshinskii-Moriya interaction. (b) External magnetic field $\bm H_{\rm ex}$ where $\theta$ is an inclination angle from the normal direction. (c), (d) Color maps of the normal component of magnetizations $m_z$ (c) and scalar spin chiralities $c_{\rm s}$ (d) of a skyrmion under a perpendicular $\bm H_{\rm ex}$ field. In-plane components of the magnetizations $(m_x,m_y)$ are displayed by arrows. (e), (f) Those under an inclined $\bm H_{\rm ex}$ field with $\theta$=30$^\circ$.}
\label{Fig1}
\end{figure}
%%%%%%%%%%%%%%%%%%%%%%%%%%%%%%%%%%%%%%%%%%%%%%%%%%%%%%%%%%%%%
A skyrmion crystal, a hexagonally crystalized state of magnetic skyrmions~\cite{Bogdanov89,Bogdanov94,Muhlbauer09,YuXZ10,Nagaosa13,Seki15}, has characteristic resonance modes at microwave frequencies~\cite{Mochizuki12,Petrova11,Schwarze15,Garst17}, which give rise to intriguing physical phenomena~\cite{Mochizuki15a} such as microwave directional dichroism~\cite{Mochizuki13,Okamura13,Mochizuki15b,Okamura15}, spin-voltage induction~\cite{Ohe13,Shimada15}, and spin-current generation~\cite{Hirobe15}. When a static magnetic field $\bm H_{\rm ex}$ is applied perpendicular to a thin-plate specimen of the skyrmion-hosting magnet, several types of spin-wave modes emerge depending on the microwave polarization~\cite{Mochizuki12}. A microwave magnetic field $\bm H^\omega$ normal to the skyrmion plane ($\bm H^\omega_\perp$) activates the so-called breathing mode in which all the skyrmions constituting the skyrmion crystal uniformly expand and shrink in an oscillatory manner. On the other hand, the $\bm H^\omega$ field within the skyrmion plane ($\bm H^\omega_\parallel$) activates two types of rotation modes with opposite rotational senses, in which cores of all the skyrmions circulate uniformly in counterclockwise and clockwise ways.

In addition to the crystallized form, skyrmions can appear as individual defects in a ferromagnetic state; such skyrmions are also expected to have peculiar collective modes~\cite{LinSZ14}. Isolated skyrmions confined in a nano-ferromagnet are potentially useful for applications~\cite{Finocchio16} to memory devices~\cite{Fert13}, magnonics devices~\cite{MaF15,MoonKW16,Mruczkiewicz16}, spin-torque oscillators~\cite{LiuRH15,ZhangS15}, and microwave sensing devices~\cite{Finocchio15}. As such, it is necessary to clarify the microwave-active eigenmodes of a single skyrmion in a ferromagnetic specimen. 

In addition, it is important to establish a way to manipulate isolated skyrmions using microwaves for their device application. As the microwave field $\bm H^\omega_\perp$ cannot activate precessions of the magnetizations when the microwave field is applied parallel to them, we can activate pure breathing-type skyrmion oscillations with $\bm H^\omega_\perp$ under a perpendicular $\bm H_{\rm ex}$ field without exciting the background ferromagnetic state. However, once the $\bm H_{\rm ex}$ field is inclined, the microwave magnetic field $\bm H^\omega_\perp$ excites huge ferromagnetic resonances, which drown out the weaker skyrmion resonances. Moreover, the microwave energy is absorbed by the sample when exciting the gigantic ferromagnetic resonances, which would inevitably result in high energy consumption and considerable temperature rise. Therefore, a technique to selectively activate an isolated skyrmion is urgently required.

In this Letter, we first theoretically show that the eigenmodes of an isolated skyrmion embedded in a ferromagnetic environment can be selectively activated with a microwave electric field $\bm E^\omega$ via oscillatory modulation of the Dzyaloshinskii-Moriya interaction (DMI) without activating ferromagnetic resonances. We then demonstrate that translational motion of the skyrmion can be driven through activating its resonance modes using this electric technique in an inclined $\bm H_{\rm ex}$ field. The latter part of the research was motivated by the recent theoretical work by Wang $et$ $al.$ that demonstrated skyrmion propagation via activating the breathing mode with a microwave $\bm H^\omega_\perp$ field~\cite{WangW15}. Our study reveals that skyrmion motion can be driven not only by the breathing mode but also by other $\bm E^\omega$-active modes. Furthermore, we find that a mode with combined clockwise and breathing oscillations can induce much faster propagation of the skyrmion than the previously studied breathing mode. Our findings pave a new way toward efficient manipulation of isolated skyrmions in nano-devices via the application of microwaves.

%%%%%%%%%%%%%%%%%%%%%%%%%%%%%%%%%%%%%%%%%%%%%%%%%%%%%%%%%%%%%
\begin{figure*}
\includegraphics[scale=1.0]{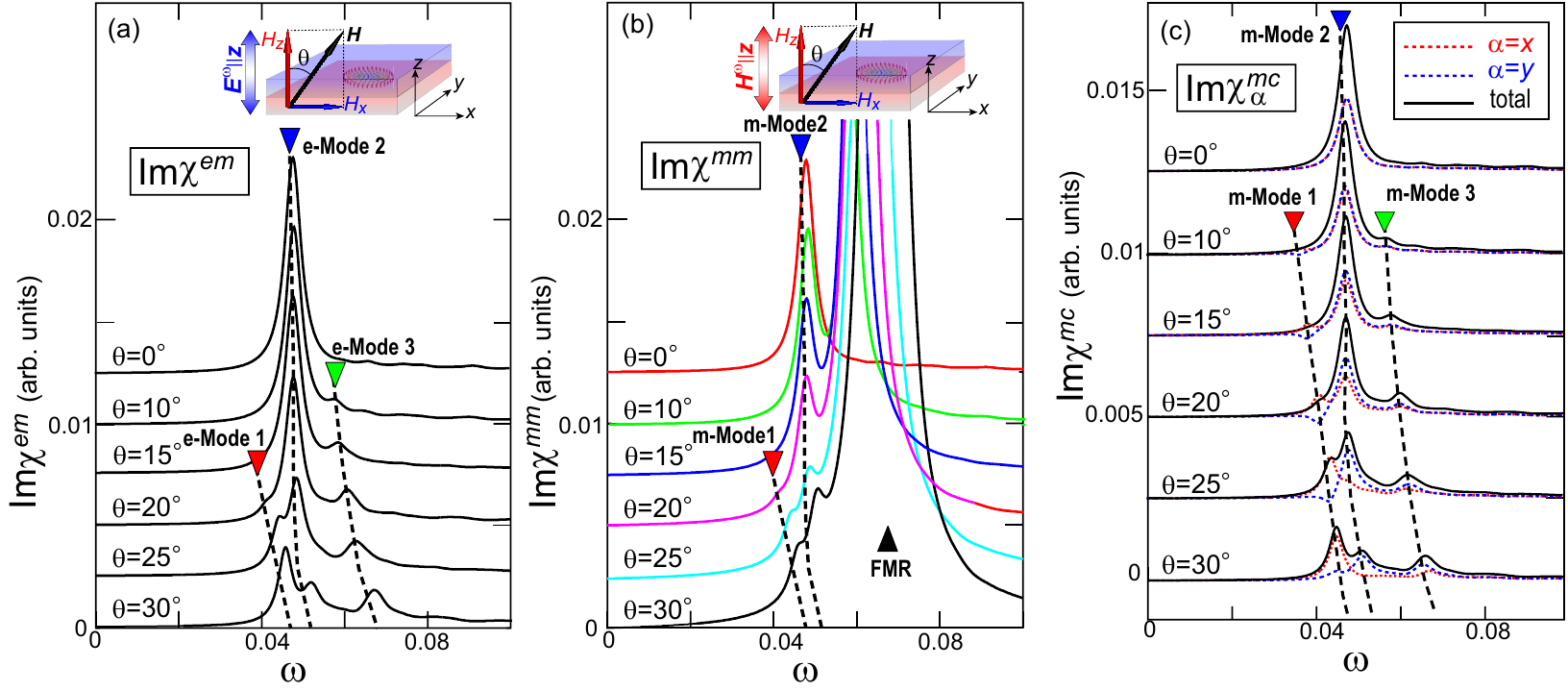}
\caption{(color online). Imaginary parts of (a) the electromagnetic susceptibility ${\rm Im} \chi^{\rm em}(\omega)$, (b) the magnetic susceptibility ${\rm Im} \chi^{\rm mm}(\omega)$, and (c) the chirality susceptibility ${\rm Im} \chi_\alpha^{\rm mc}(\omega)$ for several values of $\theta$. Here, an inclined magnetic field $\bm H_{\rm ex}=(H_z\tan\theta, 0, H_z)$ with $H_z$=0.057 is applied. Three $\bm E^\omega$-active modes are labeled as e-Modes 1-3 in (a), whereas the three $\bm H^\omega$-active modes are labeled as m-Modes 1-3 in (b) and (c). The extremely intense mode around $\omega\sim0.05-0.06$ in (b) is the ferromagnetic resonance (FMR).}
\label{Fig2}
\end{figure*}
%%%%%%%%%%%%%%%%%%%%%%%%%%%%%%%%%%%%%%%%%%%%%%%%%%%%%%%%%%%%%
We consider a magnetic bilayer system composed of a ferromagnetic layer and a heavy-metal layer with strong spin-orbit interactions [see Fig.~\ref{Fig1}(a)], where the spatial inversion symmetry is broken at their interface, and thereby the DMI is active. An inclined magnetic field $\bm H_{\rm ex}=(H_x, 0, H_z)$ with $H_x=H_z\tan\theta$ is applied where $\theta$ is the inclination angle [see Fig.~\ref{Fig1}(b)]. For $0^\circ < \theta \le 90^\circ$, the $\bm H_{\rm ex}$ field is inclined toward the positive $x$ direction. The DMI favors a rotating alignment of the magnetizations, which results in the formation of a Neel-type skyrmion. The skyrmion has a circular symmetry under a perpendicular $\bm H_{\rm ex}$ field ($\theta=0^\circ$), but has disproportionate weight in distributions of the magnetizations and scalar spin chiralities [see Fig.~\ref{Fig1}(c) and (d)]. To describe the magnetism in this magnetic bilayer system, we employ a classical Heisenberg model on a square lattice with magnetization vectors $\bm m_i$ whose norm $m$ is unity~\cite{Bak80,YiSD09}. The Hamiltonian contains the ferromagnetic-exchange interaction, the Zeeman coupling to the magnetic fields, and the interfacial DMI:
%%%%%%%%%%%%%%%%%%%%%%%%%%%%%%%%%%%
\begin{eqnarray}
& &\mathcal{H}=
-J \sum_{<i,j>} \bm m_i \cdot \bm m_j 
-[\bm H_{\rm ex}+\bm H(t)] \cdot \sum_i \bm m_i
\nonumber \\
&+&D(t) \sum_i [ (\bm m_i \times \bm m_{i+\hat{x}}) \cdot \hat{\bm y}
\;-\; (\bm m_i \times \bm m_{i+\hat{y}}) \cdot \hat{\bm x} ].
\label{eqn:model}
\end{eqnarray}
%%%%%%%%%%%%%%%%%%%%%%%%%%%%%%%%%%%
Here $\bm H(t)=(0,0,H_z(t))$ and $\bm E(t)=(0,0,E_z(t))$ represent time-dependent magnetic and electric fields applied perpendicular to the sample plane, respectively. We neglect magnetic anisotropies because they do not alter the results qualitatively although stability of the skyrmions and resonant frequencies of the eigenmodes may be slightly changed. The strength of the interfacial DMI can be tuned by applying a gate electric field normal to the plane via varying the extent of the spatial inversion asymmetry~\cite{Nawaoka15,Srivastava18}. The DMI coefficient $D(t)=D_0+\Delta D(t)$ has two components, namely, a steady component $D_0$ and a $\bm E(t)$-dependent component $\Delta D(t)=\kappa E_z(t)$ with $\kappa$ being the coupling constant. We take $J$=1 for the energy units and take $D_0/J$=0.27. For the inclined magnetic field, we take $H_z$=0.057 with $\theta$ being a variable. 
%%Note that magnetization textures considered in the present study are slowly varying in space, and thus the coupling to the background crystal structure is significantly weak. This justifies our treatment with a spin model on a simple square lattice after the coarse graining of magnetizations.
The unit conversions when $J$=1 meV are summarized in Table~\ref{tab:uconv}.
%%%%%%%%%%%%%%%%%%%%%%%%%%%%%%%%%%%%%%%
\begin{table}
\begin{tabular}{l|cc} \hline \hline
Exchange int.  & \hspace{0.5cm} $J=1$             & \hspace{0.5cm} 1 meV \\
Time           & \hspace{0.5cm} $t=1$          & \hspace{0.5cm} 0.66 ps \\
Frequency $f=\omega/2\pi$ &\hspace{0.5cm} $\omega=0.01$ & \hspace{0.5cm} 2.41 GHz\\
Magnetic field & \hspace{0.5cm} $H=1$             & \hspace{0.5cm} 8.64 T \\ 
\hline \hline
\end{tabular}
\caption{Unit conversion table when $J$=1 meV.}
\label{tab:uconv}
\end{table}
%%%%%%%%%%%%%%%%%%%%%%%%%%%%%%%%%%%%%%%

We simulate the magnetization dynamics by numerically solving the Landau-Lifshitz-Gilbert equation using the fourth-order Runge-Kutta method. The equation is given by,
%%%%%%%%%%%%%%%%%%%%%%%%%%%%%%%%%%%
\begin{equation}
\frac{d\bm m_i}{dt}=-\gamma_{\rm m} \bm m_i \times \bm H^{\rm eff}_i 
+\frac{\alpha_{\rm G}}{m} \bm m_i \times \frac{d\bm m_i}{dt}.
\label{eq:LLGEQ}
\end{equation} 
%%%%%%%%%%%%%%%%%%%%%%%%%%%%%%%%%%%
Here $\alpha_{\rm G}$(=0.04) and $\gamma_{\rm m}$ are the Gilbert-damping constant and the gyrotropic ratio, respectively. The effective field $\bm H_i^{\rm eff}$ is calculated as
%%%%%%%%%%%%%%%%%%%%%%%%%%%%%%%%%%%
%%\begin{equation}
$\bm H^{\rm eff}_i=-(1/\gamma_{\rm m}\hbar) \partial \mathcal{H}/\partial \bm m_i$.
%%\label{eq:EFFMF}
%%\end{equation}
%%%%%%%%%%%%%%%%%%%%%%%%%%%%%%%%%%%

We first calculate the dynamical electromagnetic and magnetic susceptibilities $\chi^{\rm em}$ and $\chi^{\rm mm}$:
%%%%%%%%%%%%%%%%%%%%%%%%%%%%%%%%%%
\begin{equation}
\chi^{\rm em}(\omega) =
\sqrt{\frac{\mu_0}{\epsilon_0}}
\frac{\Delta M_z^{\omega}}{E_{\rm pulse}}, \quad
\chi^{\rm mm}(\omega) =
\frac{\Delta M_z^{\omega}}{\mu_0 H_{\rm pulse}}.
\end{equation}
%%%%%%%%%%%%%%%%%%%%%%%%%%%%%%%%%%%
After applying a short pulse $H_z(t)$ or $E_z(t)$ with duration of $\Delta t$=1 in the units of $J/\hbar$, we trace time profiles of the net magnetization $M_z(t)=(1/N)\sum_i \bm m_{zi}(t)$ and $\Delta M_z(t)=M_z(t)-M_z(0)$ and obtain the Fourier transform $\Delta M_z^{\omega}$. Dividing this quantity by an amplitude of the pulse $H_{\rm pulse}$ or $E_{\rm pulse}$, we obtain these susceptibilities. The calculations are performed using a system of $N$=160$\times$160 sites with periodic boundary conditions.

Figure~\ref{Fig2}(a) displays the imaginary parts of the electromagnetic susceptibilities ${\rm Im} \chi^{\rm em}(\omega)$ for several values of $\theta$, which describe the response of the magnetizations to the $\bm E^\omega$ field. When $\theta=0^\circ$, the spectrum exhibits a single peak corresponding to the breathing mode activated by the oscillating DMI under the application of an AC electric field. When the $\bm H_{\rm ex}$ field is inclined with $\theta\ne 0^\circ$, two novel modes emerge, one with a higher and one with a lower frequency than the breathing mode. The intensities of the novel modes increase, whereas the intensity of the original breathing mode is increasingly suppressed as $\theta$ increases.

The imaginary parts of the magnetic susceptibilities ${\rm Im} \chi^{\rm mm}(\omega)$ in Fig.~\ref{Fig2}(b) describe the response of the magnetizations to the $\bm H^\omega$ field. We find that only a breathing mode (m-Mode 2) appears when the $\bm H_{\rm ex}$ field is perpendicular ($\theta=0^\circ$); however, its intensity decreases as $\theta$ increases. Remarkably, a large ferromagnetic resonance from the surrouding ferromagnetic magnetizations emerges under an inclined $\bm H_{\rm ex}$ field, whereas it is totally silent under the perpendicular $\bm H_{\rm ex}$ field. We also find a novel mode (m-Mode 1) at lower frequencies.

In reality, the skyrmion has another $\bm H^\omega$-active mode (m-Mode 3) at higher frequencies, but it is hidden behind the gigantic ferromagnetic resonance and thus cannot be seen in the spectra of ${\rm Im} \chi^{\rm mm}(\omega)$. We can see this weak response of the skyrmion to the $\bm H^\omega$ field by focusing on the vector spin chiralities, $\bm s_i=\sum_\gamma \bm m_i \times \bm m_{i+\gamma}$ ($\gamma=\hat{x}, \hat{y}$). The calculated imaginary parts of the dynamical susceptibilities ${\rm Im} \chi_\alpha^{\rm mc}(\omega)$ for the $\alpha$-component of the vector spin chirality $S_\alpha=(1/N)\sum_i s_{\alpha i}$ ($\alpha=x,y$) are shown in Fig.~\ref{Fig2}(c). We find that they coincide with the spectra of ${\rm Im} \chi^{\rm em}(\omega)$ in Fig.~\ref{Fig2}(a). These results indicate that the magnetic method using $\bm H^\omega$ cannot selectively activate the eigenmodes of an isolated skyrmion in the ferromagnetic specimen; however, the results show that the electrical method using $\bm E^\omega$ can achieve this. This electrical technique is anticipated to play a crucial role for developing future skyrmion-based devices.

%%%%%%%%%%%%%%%%%%%%%%%%%%%%%%%%%%%%%%%%%%%%%%%%%%%%%%%%%%%%
\begin{figure}
\includegraphics[scale=0.5]{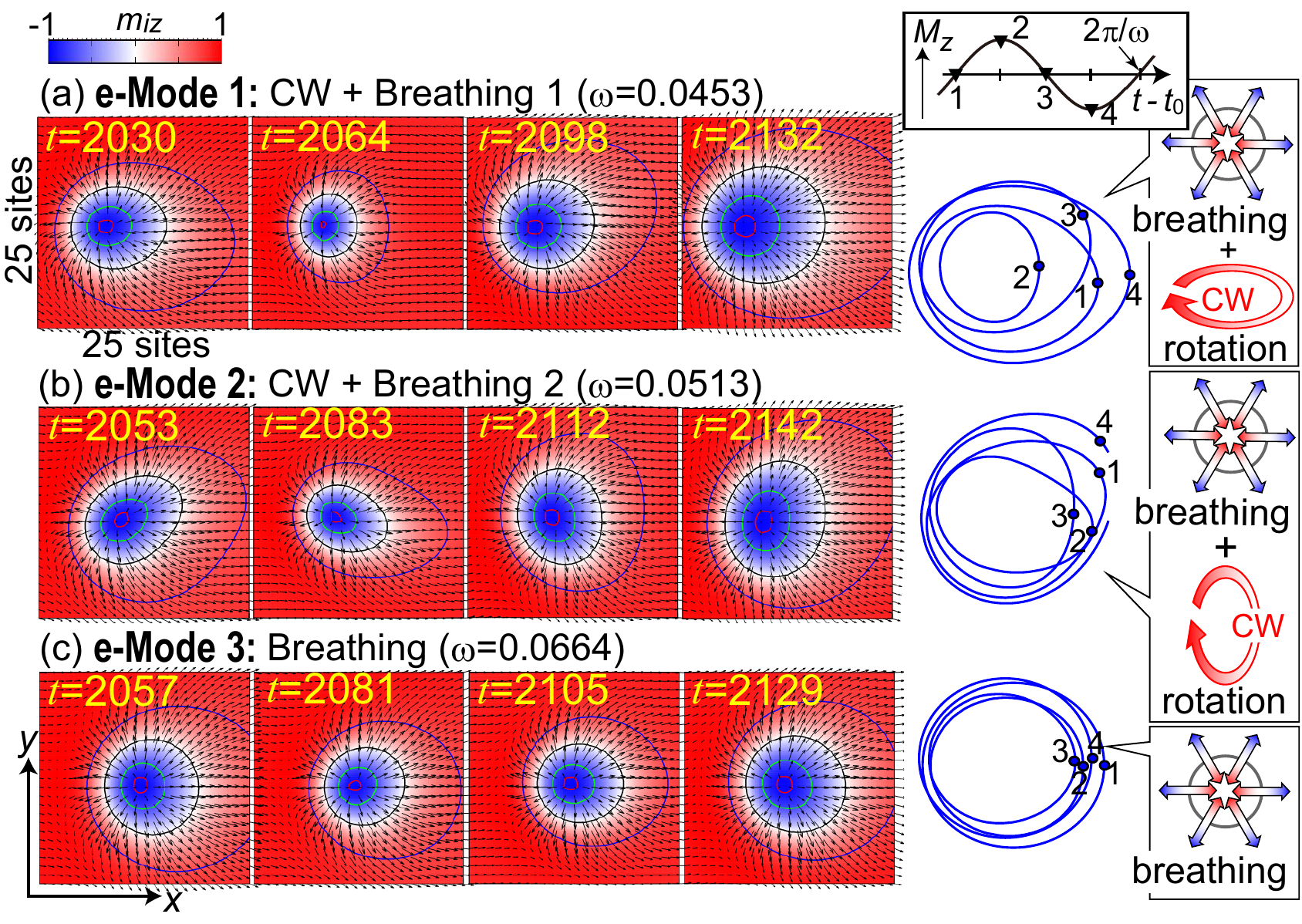}
\caption{(color online). Simulated snapshots of the magnetization dynamics for three $\bm E^\omega$-active eigenmodes (e-Modes 1-3) of an isolated skyrmion in the ferromagnetic specimen under an inclined $\bm H_{\rm ex}$ field where $H_z$=0.057 and $\theta=30^\circ$.}
\label{Fig3}
\end{figure}
%%%%%%%%%%%%%%%%%%%%%%%%%%%%%%%%%%%%%%%%%%%%%%%%%%%%%%%%%%%%%
Based on the susceptibility data, we find that an isolated skyrmion in a ferromagnetic specimen has three low-lying eigenmodes. In Fig.~\ref{Fig3}, we show simulation results of snapshots for $\theta=30^\circ$. It is found that all of these modes have the breathing component, i.e., they show oscillatory expansion and shrinkage. Among these three modes, the higher-frequency mode ($\omega$=0.0664) can be regarded as a pure breathing mode, whereas the other two modes show distinct behaviors. The lower-frequency mode ($\omega=0.0453$) is accompanied by a clockwise rotation of skyrmion in an elliptical orbit oriented horizontally against the inclined direction of $\bm H_{\rm ex}$ as shown in Fig.~\ref{Fig3}(a). The moderate-frequency mode ($\omega=0.0513$) is also accompanied by the clockwise rotation of the skyrmion in an elliptical orbit, but its trajectory is oriented vertically against the inclined direction of $\bm H_{\rm ex}$ as shown in Fig.~\ref{Fig3}(b). The higher-lying pure breathing mode at $\omega$=0.0664 does not show any rotational component [Fig.~\ref{Fig3}(c)]. The three $\bm E^\omega$-active modes are referred to as e-Modes 1, 2, and, 3 [see Fig.~\ref{Fig2}(a) and Fig~\ref{Fig3}].

%%%%%%%%%%%%%%%%%%%%%%%%%%%%%%%%%%%%%%%%%%%%%%%%%%%%%%%%%%%%%
\begin{figure}[t]
\includegraphics[scale=1.0]{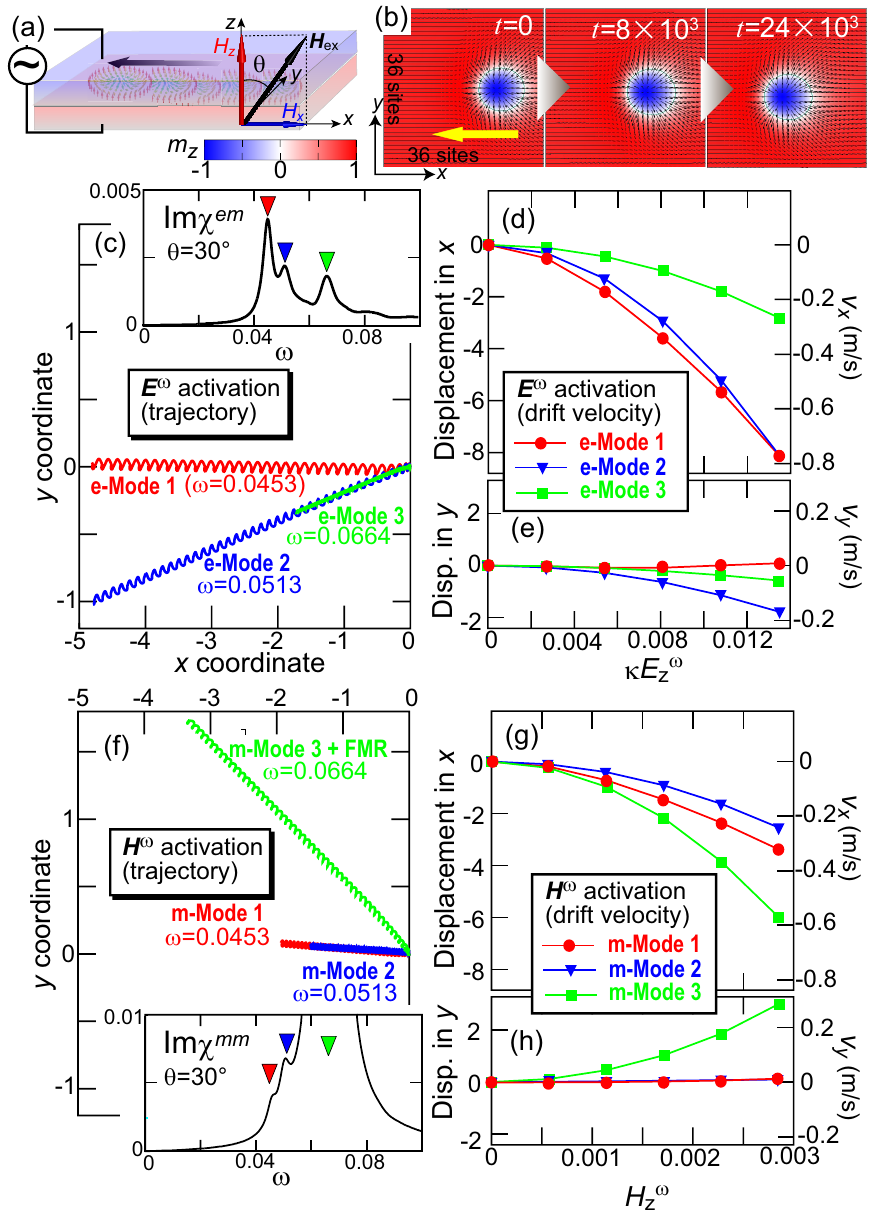}
\caption{(color online). (a) Illustration of skyrmion propagation driven by the $\bm E^\omega$ field through activating a resonant mode of the skyrmion under an inclined $\bm H_{\rm ex}$ field. (b) Simulated snapshots of the skyrmion propagation for the $\bm E^\omega$-active mode with $\omega$=0.0513 (e-Mode 2). (c)-(e) Trajectories (c) and drift velocities $v_x$ (d) and $v_y$ (e) of the propagating skyrmion driven by the $\bm E^\omega$ field for three different resonant modes. (f)-(h) Those of the propagating skyrmion driven by the $\bm H^\omega$ field. All the simulations were performed with an inclined $\bm H_{\rm ex}$ field where $H_z$=0.057 and $\theta=30^\circ$.}
\label{Fig4}
\end{figure}
%%%%%%%%%%%%%%%%%%%%%%%%%%%%%%%%%%%%%%%%%%%%%%%%%%%%%%%%%%%%%
Next we numerically investigate the translational motion of a skyrmion driven by the electrically activated resonance modes under an inclined $\bm H_{\rm ex}$ field [see Fig.~\ref{Fig4}(a)]. Here the inclination angle of $\bm H_{\rm ex}$ is fixed at $\theta=30^\circ$. The amplitude of the AC electric field $E_z(t)=E_z^\omega \sin\omega t$ is fixed at $\kappa E_z^\omega=0.05D_0=0.0135$. A recent experiment for Ta/FeCoB/TaO$_x$ reported a huge $E$-field-induced variation of the interfacial DMI that reaches 140$\%$ when the applied voltage is 10 V. This observation supports the experimental feasibility of the 5$\%$ modulation assumed here. Figure~\ref{Fig4}(b) shows simulated snapshots of the skyrmion motion when the $\bm E^\omega$ field with $\omega$=0.0513 is applied, which indeed displays propagation in the negative $x$ direction. 

The trajectories of the propagating skyrmion during a time period from $t$=0 to $t$=5000 are shown in Fig.~\ref{Fig4}(c) for three different $\bm E^\omega$-active modes at $\omega$=0.0453, 0.0513, and 0.0664. They were obtained by tracing the center-of-mass coordinate $(j_x,j_y)$ of the topological-charge distribution:
%%%%%%%%%%%%%%%%%%%%%%%%%%%%%%%%%%%%%%%%%%%%%%%%%%%%%%%%%%%%%
\begin{equation}
j_\gamma=\sum_{\bm i=(i_x,i_y)} i_\gamma c_{\rm s}(i_x,i_y)/
\sum_{\bm i=(i_x,i_y)} c_{\rm s}(i_x,i_y)
\end{equation}
%%%%%%%%%%%%%%%%%%%%%%%%%%%%%%%%%%%%%%%%%%%%%%%%%%%%%%%%%%%%%
with 
%%%%%%%%%%%%%%%%%%%%%%%%%%%%%%%%%%%%%%%%%%%%%%%%%%%%%%%%%%%%%
\begin{equation}
c_{\rm s}=\frac{1}{8\pi} \left[
\bm m_i \cdot (\bm m_{i+\hat{x}} \times \bm m_{i+\hat{y}})+
%%\right.
%%\nonumber \\
%%\left.
\bm m_i \cdot (\bm m_{i-\hat{x}} \times \bm m_{i-\hat{y}})
\right].
\end{equation}
%%%%%%%%%%%%%%%%%%%%%%%%%%%%%%%%%%%%%%%%%%%%%%%%%%%%%%%%%%%%%
%Note that origins of all the trajectories are set at $(x,y)$=(0,0). 

We find that the direction and velocity of the motion vary depending on the skyrmion resonance mode. For all the modes, the skyrmion moves mainly in the negative $x$ direction. However, the trajectories for e-Modes 2 and 3 are slightly slanted toward the negative $y$ direction; meanwhile, the trajectory for e-Mode 1 is perfectly parallel to the $x$ axis. Interestingly, despite the slanted trajectory for e-Mode 2, its traveling distance along the $x$ axis is identical to that of the trajectory for e-Mode 1, which is directed perfectly along the $x$ axis. It can also be seen that the directions of the skyrmion movement for e-Modes 2 and 3 are the same, even though their traveling distances are different. The traveling distance for e-Mode 3 is much shorter than that for e-Mode 2 because of the smaller intensity of the latter mode, as can be seen in the inset of Fig.~\ref{Fig4}(c). 

In Fig.~\ref{Fig4}(d) and (e), we plot the velocities $\bm v$=$(v_x, v_y)$ of the skyrmion for three different $\bm E^\omega$-active modes (see the right vertical axes) as functions of the strength of the time-dependent DMI, i.e., $\kappa E_z^\omega$. The values are calculated from the simulated displacements of the skyrmion in the $x$ and $y$ directions for a time period from $t$=2000 to $t$=10000 (see the left vertical axes) by assuming $J$= 1 meV and $a$=5 {\AA} with $a$ being the lattice constant. It can be seen that the velocities are proportional to the square of $\kappa E_z^\omega$, and they are on the order of 10$^{-1}$ m/s. In fact, the traveling speed of the skyrmions achieved using this technique turns out to be relatively slow compared with the speed achieved in techniques based on electric-current injection~\cite{Jonietz10,YuXZ12,Iwasaki13a,Iwasaki13b,Buttner15,WooS16,YuG17,JiangW17}. However, the present technique has an advantage: it is free from the Joule heating, and thus the energy consumption and the temperature rise could be significantly suppressed.

We finally study the skyrmion motion driven by AC magnetic fields $\bm H^\omega$ under an inclined $\bm H_{\rm ex}$ field with $H_z$=0.057 and $\theta=30^\circ$. The $\bm H^\omega$ is applied perpendicular to the skyrmion plane, which is given by $H^\omega_z \sin\omega t$ with $H^\omega_z=0.05H_z=0.00285$. The simulated trajectories and velocities $\bm v$=$(v_x, v_y)$ are shown in Fig.~\ref{Fig4}(f) and (g), respectively. We find that the trajectories are again oriented almost in the negative $x$ direction; however, for the ferromagnetic resonance mode with $\omega$=0.0664 the trajectory is slanted toward the positive $y$ direction, which contrasts with the case of the $\bm E^\omega$-active mode. Noticeably, the higher-frequency mode with $\omega$=0.0664 has a much faster propagation of the skyrmion than the other two modes. However, the usage of this mode is not energetically efficient because this mode is not an eigenmode of the isolated skyrmion but a very intense resonance of the vast ferromagnetic magnetizations, which necessarily leads to large energy consumption and considerable rise of device temperatures. 
%%Therefore, we can conclude that the proposed technique based on the selective activation of the $\bm E^\omega$-active mode is advantageous for efficient manipulation of isolated skyrmions for future device applications.

In summary, we have theoretically found that resonance modes of an isolated sykrmion in a ferromagnet can be activated by application of AC electric fields through oscillatory variation of the interfacial DMI. The advantage of this method compared with conventional methods using AC magnetic fields is that we can selectively excite skyrmions without activating gigantic ferromagnetic resonances; this results in a significant suppression of both energy consumption and temperature rise. Our result revealed that among the three $\bm E^\omega$-active modes, the mode with combined clockwise and breathing oscillations induces much faster skyrmion propagation than the previously studied breathing mode. Our findings will pave a new way toward the efficient manipulation of isolated skyrmions and thus will be useful for future skyrmion-based devices.

This work was supported by JSPS KAKENHI (Grant No. 17H02924), Waseda University Grant for Special Re-search Projects (Project Nos. 2017S-101, 2018K-257), and JST PRESTO (Grant No. JPMJPR132A).
%%%%%%

\end{document}